\pdfoutput=1

\documentclass[11pt,twoside,a4paper,cmspaper,final,collab]{cms-tdr}
\def\svnVersion{415779}\def\cmsCernNoTag{CERN-EP/2017-023}\def\cmsCernDate{\today}\def\cmsMessage{Published in Physics Letters B as \href{http://dx.doi.org/10.1016/j.physletb.2017.06.064}{\doi{10.1016/j.physletb.2017.06.064.}}}

\begin{document}\cmsNoteHeader{TOP-16-016}

\hyphenation{had-ron-i-za-tion}
\hyphenation{cal-or-i-me-ter}
\hyphenation{de-vices}
\RCS$Revision: 412701 $
\RCS$HeadURL: svn+ssh://svn.cern.ch/reps/tdr2/papers/TOP-16-016/trunk/TOP-16-016.tex $
\RCS$Id: TOP-16-016.tex 412701 2017-06-24 18:46:45Z alverson $
\newlength\cmsFigWidth
\ifthenelse{\boolean{cms@external}}{\setlength\cmsFigWidth{0.85\columnwidth}}{\setlength\cmsFigWidth{0.4\textwidth}}
\ifthenelse{\boolean{cms@external}}{\providecommand{\cmsFirst}{top\xspace}}{\providecommand{\cmsFirst}{upper left\xspace}}
\ifthenelse{\boolean{cms@external}}{\providecommand{\cmsMiddle}{middle\xspace}}{\providecommand{\cmsMiddle}{upper right\xspace}}

\newcommand{\tttt}{\ensuremath{\ttbar\ttbar}\xspace}
\newcommand{\njets}{\ensuremath{N_{\mathrm{j}}}\xspace}
\newcommand{\njetsw}{\ensuremath{N_{\mathrm{j}}^{\mathrm{w}}}\xspace}
\newcommand{\nltags}{\ensuremath{N_{\text{tags}}^{\mathrm{l}}}\xspace}
\newcommand{\nmtags}{\ensuremath{N_{\text{tags}}^{\mathrm{m}}}\xspace}
\newcommand{\htb}{\ensuremath{H_{\mathrm{T}}^{\mathrm{b}}}\xspace}
\newcommand{\htrat}{\ensuremath{H_{\mathrm{T}}^{\text{ratio}}}\xspace}
\newcommand{\httwom}{\ensuremath{H_{\mathrm{T}}^{\mathrm{2m}}}\xspace}
\newcommand{\thirdjetpt}{\ensuremath{p_{\mathrm{T}}^{\mathrm{j3}}}\xspace}
\newcommand{\fourthjetpt}{\ensuremath{p_{\mathrm{T}}^{\mathrm{j4}}}\xspace}
\newcommand{\leadleppt}{\ensuremath{\pt^{\ell 1}}\xspace}
\newcommand{\eventsph}{\ensuremath{S}\xspace}
\newcommand{\hth}{\ensuremath{C}\xspace}
\newcommand{\drll}{\ensuremath{\Delta R_{\ell\ell}}\xspace}
\newcommand{\drbb}{\ensuremath{\Delta R_{\PQb\PQb}}\xspace}
\newcommand{\leadlepeta}{\ensuremath{\eta^{\ell 1}}\xspace}
\newcommand{\redhadmass}{\ensuremath{M_{\text{red}}^{\mathrm{h}}}\xspace}
\newcommand{\ourLumi}{\ensuremath{2.6\fbinv}\xspace}
\newcommand{\sigmattttSM}{\ensuremath{\sigma_{\tttt}^{\mathrm{SM}}}\xspace}
\newcommand{\BDTtrijetone}{\ensuremath{T_{\text{trijet1}}}\xspace}
\newcommand{\BDTtrijettwo}{\ensuremath{T_{\text{trijet2}}}\xspace}
\newcommand{\BDTdilep}{\ensuremath{D_{\tttt}^{\text{dil}}}\xspace}
\newcommand{\BDTljets}{\ensuremath{D_{\tttt}^{\mathrm{lj}}}\xspace}
\newcommand{\htx}{\ensuremath{H_{\mathrm{T}}^{\mathrm{x}}}\xspace}
\newcommand{\irel}{\ensuremath{I_\text{rel}}\xspace}
\newcommand{\xsecmusinglepton}{\ensuremath{17.2}}
\newcommand{\xsecmusingleptonexp}{\ensuremath{16.4}}
\newcommand{\xsecmusingleptonup}{\ensuremath{9.8}}
\newcommand{\xsecmusingleptondown}{\ensuremath{5.7}}
\newcommand{\xsecfbsinglepton}{\ensuremath{158}}
\newcommand{\xsecfbsingleptonexp}{\ensuremath{151}}
\newcommand{\xsecfbsingleptonup}{\ensuremath{90}}
\newcommand{\xsecfbsingleptondown}{\ensuremath{52}}
\newcommand{\xsecmudilepton}{\ensuremath{14.5}}
\newcommand{\xsecmudileptonexp}{\ensuremath{24.7}}
\newcommand{\xsecmudileptonup}{\ensuremath{16.7}}
\newcommand{\xsecmudileptondown}{\ensuremath{9.2}}
\newcommand{\xsecfbdilepton}{\ensuremath{134}}
\newcommand{\xsecfbdileptonexp}{\ensuremath{227}}
\newcommand{\xsecfbdileptonup}{\ensuremath{154}}
\newcommand{\xsecfbdileptondown}{\ensuremath{84}}
\newcommand{\xsecmucombo}{\ensuremath{10.2}}
\newcommand{\xsecmucomboexp}{\ensuremath{12.8}}
\newcommand{\xsecmucomboup}{\ensuremath{8.3}}
\newcommand{\xsecmucombodown}{\ensuremath{4.5}}
\newcommand{\xsecfbcombo}{\ensuremath{94}}
\newcommand{\xsecfbcomboexp}{\ensuremath{118}}
\newcommand{\xsecfbcomboup}{\ensuremath{76}}
\newcommand{\xsecfbcombodown}{\ensuremath{41}}
\newcommand{\xsecmucomboall}{\ensuremath{7.4}}
\newcommand{\xsecmucomboallexp}{\ensuremath{7.7}}
\newcommand{\xsecmucomboallup}{\ensuremath{4.1}}
\newcommand{\xsecmucomboalldown}{\ensuremath{2.6}}
\newcommand{\xsecfbcomboall}{\ensuremath{69}}
\newcommand{\xsecfbcomboallexp}{\ensuremath{71}}
\newcommand{\xsecfbcomboallup}{\ensuremath{38}}
\newcommand{\xsecfbcomboalldown}{\ensuremath{24}}
\newcommand{\ttttPredThirteen}{\ensuremath{9.2^{+2.9}_{-2.4}}\xspace}
\newcommand{\T}{\rule{0pt}{2.6ex}}
\newcommand{\B}{\rule[-1.2ex]{0pt}{0pt}}

\cmsNoteHeader{TOP-16-016}
\title{Search for standard model production of four top quarks in proton-proton collisions at $\sqrt{s}= 13\TeV$}

\date{\today}

\abstract{
A search for events containing four top quarks (\tttt) is reported from
proton-proton collisions recorded by the CMS experiment at
$\sqrt{s} = 13$\TeV and corresponding to an integrated
luminosity of 2.6\fbinv.
The analysis considers the single-lepton (e or $\mu$)+jets
and the opposite-sign dilepton ($\PGmp\PGmm$, $\PGm^{\pm} \Pe^{\mp}$, or $\Pep\Pem$)+jets channels.
It uses boosted decision trees to combine information on the global event and jet properties to distinguish between \tttt and \ttbar production.
The number of events observed after all selection requirements is consistent with expectations from background and standard model signal predictions, and an upper limit is set on the
cross section for \tttt production in the standard model of 94\unit{fb}  at 95\% confidence level
($10.2~\times$ the prediction), with an expected limit of 118\unit{fb}.
This is combined with the results from the published CMS search in the same-sign dilepton channel, resulting in an improved limit of 69\unit{fb} at 95\% confidence level ($7.4~\times$ the prediction), with an expected limit of 71\unit{fb}.
These are the strongest constraints on the rate of \tttt production to date.
}

\hypersetup{
pdfauthor={CMS Collaboration},
pdftitle={Search for standard model production of four top quarks in proton-proton collisions at sqrt(s) = 13 TeV},
pdfsubject={CMS},
pdfkeywords={CMS, physics, top, BSM}}

\maketitle

\section{Introduction}

\begin{figure}[h!tb]
\centering
\includegraphics[width=0.45\textwidth]{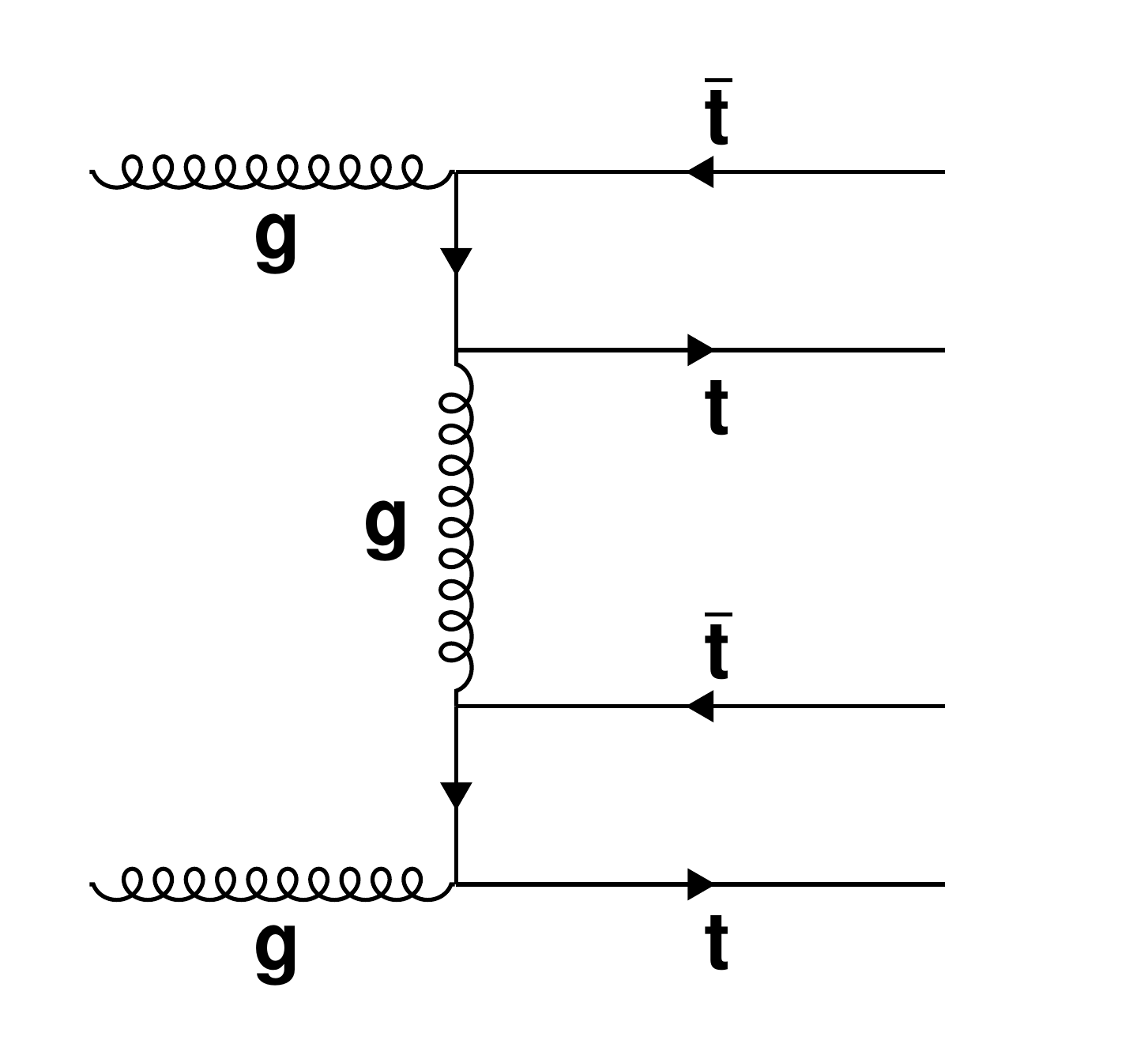}
\caption{A representative Feynman diagram for \tttt production in the SM at lowest order in QCD.}
\label{fig:tttt_SM}
\end{figure}

In proton-proton collisions at the CERN LHC most top quarks are produced in \ttbar pairs, with a small contribution from single top quark production.
It is also possible, however,
to produce four top quarks (\tttt) in the standard model (SM) via higher-order diagrams in quantum chromodynamics (QCD), mainly via gluon fusion, as shown in Fig.~\ref{fig:tttt_SM}.

The SM cross section for \tttt production is approximately five orders of magnitude smaller than that for \ttbar, and \tttt production has not yet been observed.
Observing \tttt production consistent with predictions would provide a valuable test of higher-order perturbative QCD calculations.
In addition, many models of physics beyond the SM predict an increase in the \tttt cross section owing either to the presence of hypothetical particles that decay into top quarks or to modified couplings.
These include models with massive colored bosons, Higgs boson and top quark compositeness, or extra dimensions, models with extended scalar sector such as 2HDM models~\cite{hdmrev}, and supersymmetric extensions of the SM~\cite{Toharia:2005gm,Kumar:2009vs,Calvet:2012rk,Cacciapaglia:2015eqa,Ducu:2015fda,Arina:2016cqj,Cao:2016wib}.
Some of these models predict enhancements in the observed \tttt cross section, with the associated kinematic distributions remaining similar to those from SM production. This is particularly the case when no new particles beyond those in the SM are produced on-shell.

At $\sqrt{s}=8$\TeV, where the SM predicts a cross section ($\sigmattttSM$) of 1.3\unit{fb}~\cite{Bevilacqua:2012em}, CMS set a 95\% confidence level (CL) upper limit of 32\unit{fb} on the production cross section, and comparable results, 23\unit{fb}, were obtained by ATLAS~\cite{Khachatryan2014,Aad:2015kqa}.
At $\sqrt{s}=13$\TeV, the SM prediction increases to 9.2\unit{fb}~\cite{Alwall:2014hca,Bevilacqua:2012em}, where CMS set a 95\% CL upper limit of 119\unit{fb} using a same-sign dilepton analysis~\cite{Khachatryan:2016kod}. The studies presented in this paper are performed in two separate decay modes that are complementary to and statistically independent from the same-sign dilepton analysis. The first analysis examines the final state where only one of the four W bosons from the top quark decays in \tttt production decays to a muon or electron. This single-lepton final state has the largest branching fraction in \tttt production. The second analysis focuses on the opposite-sign dilepton channel with exactly two of any combination of electrons or muons.
Both use the 13\TeV data recorded by the CMS experiment in 2015 (corresponding to an integrated luminosity of 2.6 \fbinv), and apply multivariate techniques to discriminate between the \tttt and \ttbar processes. In order to enhance the sensitivity, the search is performed in multiple jet and b jet multiplicity categories.
\section{The CMS detector}
The central feature of the CMS detector is a superconducting solenoid of 6 m internal diameter, providing a magnetic field of 3.8\unit{T}. A silicon pixel and strip tracker, a lead tungstate crystal electromagnetic calorimeter, and a brass and scintillator hadron calorimeter, each composed of a barrel and two endcap sections, are located within the solenoid volume. Muons are measured in gas-ionization detectors embedded in the steel flux-return yoke outside the solenoid. A two-tier trigger system selects the relevant collisions for offline analysis \cite{Khachatryan:2016bia}. A more detailed description of the CMS detector, together with a definition of its coordinate system and kinematic variables, can be found in \cite{CMSdet}.

\section{Data and Simulation\label{sec:datasim}}

Several Monte Carlo (MC) generators are used to simulate the signal and background processes.
The \tttt signal is simulated using the next-to-leading-order (NLO) \textsc{mg}5\_a\MCATNLO
generator (v2.2.2)~\cite{Alwall:2014hca,Mangano:2006rw}, assuming a top quark mass ($m_{\PQt}$) of 172.5\GeV.
The \tttt cross section is calculated to be \ttttPredThirteen\unit{fb}~\cite{Alwall:2014hca}, where the uncertainty includes contributions from the factorization and renormalization scale uncertainties and the dependence on the choice of parton distribution functions (PDFs).

The dominant background process is \ttbar production,
which is simulated at NLO using  \POWHEG v2 \cite{Nason:2004rx,Frixione:2007vw,Alioli:2010xd,Alioli2012}.
The events coming from Drell--Yan ($\qqbar\to \Z/\gamma^*\to\ell^+\ell^-$, with $\ell=\Pe$ or $\PGm$)+jets
and W boson+jets production are modeled using \MADGRAPH~\cite{Alwall:2014hca}, with
the MLM matching scheme~\cite{Alwall:2008}, and up to three jets.  Single top quark production and the production of \ttbar pairs in conjunction with a Higgs boson give small background contributions, and these are simulated using \POWHEG~v1.
Lastly, the production of a \ttbar pair in association with a W or Z boson
is modeled using the \textsc{mg5}\_a\MCATNLO  generator.
In all of the simulations, the initial- and final-state radiation (ISR and FSR), and the fragmentation and hadronization of quarks are modeled using \PYTHIA 8.212~\cite{1126-6708-2006-05-026,Sjostrand2008852} with the underlying event tune CUETP8M1 \cite{Khachatryan:2015pea}.
This tune uses a value of 0.137 for the strong coupling $\alpha_S(M_{\Z})$ in the parton shower simulation, which leads to a mismodeling of the jet multiplicity (\njets) spectrum. All of the simulations are corrected for this effect using factors that depend on the number of observed jets and which are equal to the ratio of the particle-level cross sections calculated assuming $\alpha_S(M_{\Z})=0.137$ and $\alpha_S(M_{\Z})=0.113$. The latter value was derived from a comparison of the predictions of the \ttbar simulation and a CMS measurement on an independent \ttbar dataset at 8\TeV \cite{Khachatryan:2015mva}.

Detailed studies show that the contributions of \ttbar{}+H, Z boson, or W boson do not significantly affect the sensitivity of the analysis. This combined background is included in the overwhelming \ttbar background for the rest of this paper, unless mentioned otherwise.  Backgrounds containing Z and W bosons but no top quarks are further referred to as electroweak (EW) backgrounds.  Among single top production modes, the only relevant one is the contribution from tW production. All of these backgrounds are included in the analysis but are orders of magnitude smaller than the background originating from \ttbar{}+jets production.

The NNPDF 3.0 and NNPDF\_nlo\_as\_0118~\cite{Ball:2011uy} PDFs are used to generate all events,
the latter being used for samples created with NLO generators.
The simulated samples already include an estimate of the additional pp interactions per bunch crossing (pileup), and further corrections are applied to make the simulation of the number of additional interactions representative of that observed in the data.
All of the simulated events are propagated through a simulation of the CMS detector, which is based on \GEANTfour (v.9.4)~\cite{Agostinelli2003250}.
The \ttbar background process is normalized to the next-to-next-to-leading-order cross section~\cite{Czakon:2011xx}.
In all other cases, the NLO cross sections are used \cite{Aliev:2010zk,Kant:2014oha,Campbell:2006wx}.

\section{Event selection}

The final states considered in this analysis are the single-lepton channel with exactly one muon or electron, and the opposite-sign dilepton channel with exactly $\PGmp\PGmm$, $\PGm^{\pm} \Pe^{\mp}$,  or $\Pep\Pem$.
These leptons originate from the W bosons from top quark decays and tend to be isolated from jets, unlike leptons produced in the decay of
B or other hadrons within jets.
Single-lepton events were recorded using a trigger that required at least one isolated muon with $\pt>18$\GeV
or one isolated electron with $\pt> 23\GeV$.
Dilepton events were recorded using a trigger that required an electron or muon with  $\pt> 17$\GeV, in combination with a second lepton where the requirement is $\pt> 8\GeV$ for a muon and 12\GeV for an electron.

Each event is required to have at least one reconstructed vertex. The primary vertex is chosen as the one with the largest value of $\sum{\pt^2}$ of the tracks associated with it.
Single-lepton events are required to contain exactly one isolated muon or electron
with $\pt> 26\GeV$ or $30\GeV$, respectively,
and pseudorapidity $\abs{\eta}<2.1$. The isolation is ensured by demanding the variable \irel to be below the predefined threshold. The relative isolation, \irel, is defined as the scalar \pt sum of the additional particles emanating from the same vertex as the lepton, within a cone of angular radius $\Delta R = \sqrt{\smash[b]{(\Delta\eta)^{2} + (\Delta\phi)^{2}}}$ = 0.4  around the lepton, divided by the \pt of the lepton, where $\Delta\eta$ and $\Delta\phi$ are the differences in pseudorapidity and azimuthal angle (in radians), respectively, between the directions of the lepton and the additional particle. The sum does not include the \pt of the muon and is corrected for the neutral particle contribution from pileup on an event-by-event basis. Electron candidates are required to satisfy restrictive identification criteria, including isolation, which are described in Ref.~\cite{Khachatryan:2015hwa}. Muons are required to satisfy the criteria described in Ref.~\cite{CMS-PAPER-MUO-10-004} and have a relative isolation variable, \irel, smaller than 0.15.

In the dilepton channel, events are required to contain two isolated leptons of opposite charge
with $\pt >20\GeV$ or 25\GeV for muons or electrons, respectively, and $\abs{\eta} < 2.4$.
Electron candidates are required to satisfy the same identification criteria as in the single-lepton channel. Because of the lower background,
the muon isolation requirement is relaxed to $\irel< 0.25$.
In the $\mu \mu$ and ee channels, the lepton pair is also required to have an invariant mass greater than 20\GeV
and outside of a 30\GeV window centered on the Z boson mass, to exclude leptons from the decays of low-mass resonances and Z bosons.
Events containing additional isolated charged leptons are vetoed.

Jets are reconstructed using a particle-flow algorithm in which the particles are clustered using the anti-\kt algorithm~\cite{Cacciari:2008gp,FastJet} with a distance parameter of 0.4. The jet momentum is determined from the vectorial sum of all particle momenta in the jet, and is reconstructed to within 5--10\% of the true momentum over the full range of \pt within the detector acceptance, as determined from simulations. An offset correction is applied to jet energies to take into account the contribution from pileup. Jet energy corrections are derived for simulation, and verified using in situ measurements of the energy balance in dijet and photon+jet events. These corrections are applied as a function of the jet \pt and $\eta$ to both data and simulated events~\cite{Chatrchyan:2011ds}.

A minimum of six or four jets are each required to have $\pt > 30$\GeV and $\abs{\eta} <2.5$ in the single-lepton or dilepton channel,
respectively, with two or more required to be tagged as originating from the hadronization of b quarks (b jets) using the combined secondary vertex (CSV) algorithm~\cite{CMS:2015csvv2}.
A working point (``medium") of the algorithm is chosen to give a misidentification rate of approximately 1\% for light-quark and gluon jets, with b tagging efficiencies of 40--75\% depending on the kinematic properties of the jet.
In the dilepton channel, events are also required to have $\HT> 500$\GeV, where \HT is defined as the scalar sum of the \pt of all jets.
This selection removes a large amount of the \ttbar background, while not significantly reducing the expected number of signal events.

The efficiency of the lepton selection is measured using tag-and-probe techniques \cite{CMS:2012llxs,CMS:2011WZ}. The simulation is corrected using \pt- and $\eta$-dependent
scale factors of order unity to provide consistency with the data.
The distribution of the CSV discriminant in simulation is corrected to match that observed in data~\cite{CMS:2013b,Khachatryan:2014qaa}.
The relative rates of \ttbar{}+\bbbar and \ttbar{}+light-quark events in the simulation are corrected to make them consistent with previous measurements~\cite{CMS:2016ttbb}.
After implementing the complete event selection criteria and all corrections, good consistency in kinematic distributions is found between the data and simulation in both channels.
The total number of events selected in the single-electron (muon) channel is 3740\,(5600) and 332 events are selected in the dilepton channel. All are consistent with the expected background. Selected events are predominantly (over 97\% in the single-lepton channel and 95\% in the dilepton channel) expected to be \ttbar{}+jets, with the remaining fraction from single top, Drell--Yan+jets, and W+jets production.

\section{Multivariate analysis}

Two boosted decision trees (BDTs), implemented using the TMVA library~\cite{bdt1,bdt2,tmva}, are used to improve
 the discrimination between signal and background. The principal differences between \tttt and \ttbar production
are
in the jet multiplicity and the number of b jets. These and the associated kinematic variables feature strongly in the choice of BDT input parameters.
This is based on the strategy developed for the CMS $\sqrt{s}=8$\TeV analysis, and the training uses simulated backgrounds from \ttbar{}+jets and \tttt production~\cite{Khachatryan2014}.

The first BDT is used to identify combinations of three jets (trijet) consistent with being the product of all-hadronic top quark decays, rather
than of other sources such as ISR or FSR.
The BDT uses the invariant dijet and trijet masses, b tagging information, and the angles between the three jets as input variables.
All possible trijet permutations are ranked according to their BDT discriminant value, from highest to lowest.
In the dilepton channel, the \ttbar background contains no hadronic top decays, so the BDT output for the
first-ranked trijet (\BDTtrijetone) is used as the discriminant. In the single-lepton channel, each background event contains a genuine hadronic top quark decay, so the jets included in \BDTtrijetone are removed and the highest-ranked BDT discriminant using the remaining jets (\BDTtrijettwo) is used.

The second BDT, with discriminants \BDTljets for the single-lepton channel and  \BDTdilep for the dilepton channel, takes the discriminant from the trijet associations as one of its input parameters.
Additional inputs are then optimized separately for the two channels and are based on the characteristics of the lepton and jet activity in the events. Not all inputs are used by both analyses. These are grouped into three categories: event activity, event topology, and b quark multiplicity. Although many of these are correlated, each one contributes some additional discrimination between the \ttbar background and the \tttt signal.

Comparison of simulated \ttbar and \tttt events leads to the selection of the following event activity variables:
\begin{enumerate}

\item The number of jets present in the event, \njets.

\item Weighted jet multiplicity (\njetsw), based on both the jet multiplicity and the \pt distribution of the jets. This quantity is sensitive to the differences between the \pt spectra of the jets from top quark decays and those originating from gluon radiation, having  higher values in events with many high-\pt jets than in events where only a few jets have high \pt and the rest are close to the selection threshold.
	It is defined as
\ifthenelse{\boolean{cms@external}}{
\begin{equation}
\begin{aligned}
\njetsw &= \frac{\int_{30}^{125}{\njets\left(\pt > \pt^\text{th}\right)\, \pt^\text{th}\,\rd\pt^\text{th}}} {\int_{30}^{125}{\pt^\text{th}\,\rd\pt^\text{th}}}\\
& = \frac{1}{14725\GeV^2}\sum_{i=0}^{\njets}{\njets\left(\pt>\pt^{i}\right)\,\left.\left(\pt^\text{th}\right)^2\right|^{\pt^{i+1}}_{\pt^{i}}}
\label{eqn:WjetMul}
\end{aligned}
\end{equation}}{
\begin{equation}
\njetsw = \frac{\int_{30}^{125}{\njets\left(\pt > \pt^\text{th}\right)\, \pt^\text{th}\,\rd\pt^\text{th}}} {\int_{30}^{125}{\pt^\text{th}\,\rd\pt^\text{th}}} = \frac{1}{14725\GeV^2}\sum_{i=0}^{\njets}{\njets\left(\pt>\pt^{i}\right)\,\left.\left(\pt^\text{th}\right)^2\right|^{\pt^{i+1}}_{\pt^{i}}}
\label{eqn:WjetMul}
\end{equation}
}
where the limits of integration are in \GeV and $\njets\left(\pt > \pt^\text{th}\right)$ is the number of jets with \pt above a threshold $\pt^\text{th}$, while $\pt^0= 30$\GeV, $\pt^{i}\,(\pt^{i+1}\ge\pt^{i})$ is the $\pt$ of the $i$th jet and $\pt^{\njets+1}=125$\GeV. The lower limit of 30\GeV is driven by the minimum \pt requirements on the jets, while the upper limit of 125\GeV is chosen since above this value, there are few events and increasing the upper limit on $\pt^\text{th}$ would not significantly affect the sensitivity of \njetsw.

\item The variable \htb, defined as the scalar sum of the \pt of all jets that are identified as b jets by the CSV algorithm, applied at its medium working point.

\item The ratio (\htrat) of the \HT of the four highest-\pt jets in the event in the single-lepton, or the two highest-\pt jets
in the dilepton channel, to the \HT of the other jets in the event.

\item The quantity \httwom, defined as the \HT in the event minus the scalar sum of the \pt of the two highest-\pt b jets.

\item The transverse momenta of the jets with the third- and fourth-largest \pt in the event (\thirdjetpt and \fourthjetpt).

\item The reduced event mass (\redhadmass), defined as the invariant mass of the system comprising all the jets in the reduced event,
where the reduced event is constructed by subtracting the jets contained in \BDTtrijetone in single-lepton events.
In \ttbar events, the reduced event will typically only contain the b jet from the semileptonic top quark decay and jets arising from ISR and FSR. Conversely, a reduced \tttt event can contain up to two hadronic top quarks and, as a result, numerous energetic jets.

\item The reduced event \HT (\htx). This is defined as the \HT of all jets in the single-lepton event selection excluding those contained in \BDTtrijetone .

\end{enumerate}

The event topology is characterized by the variables:
\begin{enumerate}

\item Event sphericity ($S$)~\cite{sphericity}, calculated from all of the jets in the event in terms of the tensor
\begin{linenomath}
$S^{\alpha \beta} = {\sum_{i} p_i^\alpha p_i^\beta}/{\sum_i \abs{\vec{p_i}}^2} \label{eq:sphTensor},$
\end{linenomath}
where $\alpha$ and $\beta$ refer to the three-components of the momentum of the $i$th  jet. The sphericity is then $S = (3/2)(\lambda_2+\lambda_3)$, where $\lambda_2$ and $\lambda_3$ are the two smallest eigenvalues of $S^{\alpha \beta}$.
The sphericity in \tttt events should differ from that in background \ttbar events of the same energy, since the jets in \ttbar events will be less isotropically distributed because of their recoil from sources such as ISR.

\item Hadronic centrality ($C$), defined as the value of \HT divided by the sum of the energies of all jets in the event.
\end{enumerate}

Since all the previous variables rely only on the hadronic information in the event,
sensitivity to the lepton information is provided through the \pt and $\eta$ of the
highest-\pt lepton (or only lepton for the single-lepton channel) $\left(\leadleppt,\leadlepeta\right)$ and the angular difference (\drll) between the leptons in dilepton events.
Finally, the b jet multiplicity is characterized in terms of the number of b jets tagged by the CSV algorithm operating
at its loose~\cite{CMS:2013b} (\nltags) and medium (\nmtags) operating points, and the angular separation \drbb between the b-tagged jets with the highest CSV discriminants, and the third- and fourth-highest discriminant values.

The event-level discriminants \BDTljets and \BDTdilep are optimized separately, resulting in the choice of different sets of variables.
For the single-lepton channel, the optimal variable set, in order of sensitivity, is found to be \njets, \BDTtrijettwo, \htb, \htrat, \leadleppt, \njetsw, \redhadmass, \htx, and the third- and fourth-highest CSV discriminants.
In the dilepton channel, the optimal variable set, in order of sensitivity, is
\njets, \njetsw, \eventsph, \BDTtrijetone, \nltags, \nmtags, \drbb, \htb, \leadleppt, \leadlepeta, \htrat, \httwom, \drll, \hth, \thirdjetpt, and \fourthjetpt. The MC modeling of the individual observables utilised in the discriminants \BDTljets and \BDTdilep was verified using control samples of \ttbar events and found to be in agreement with the data in all the jet multiplicities and b tag multiplicities.

\begin{figure*}[ht!]
\centering
    \includegraphics[width=0.45\textwidth]{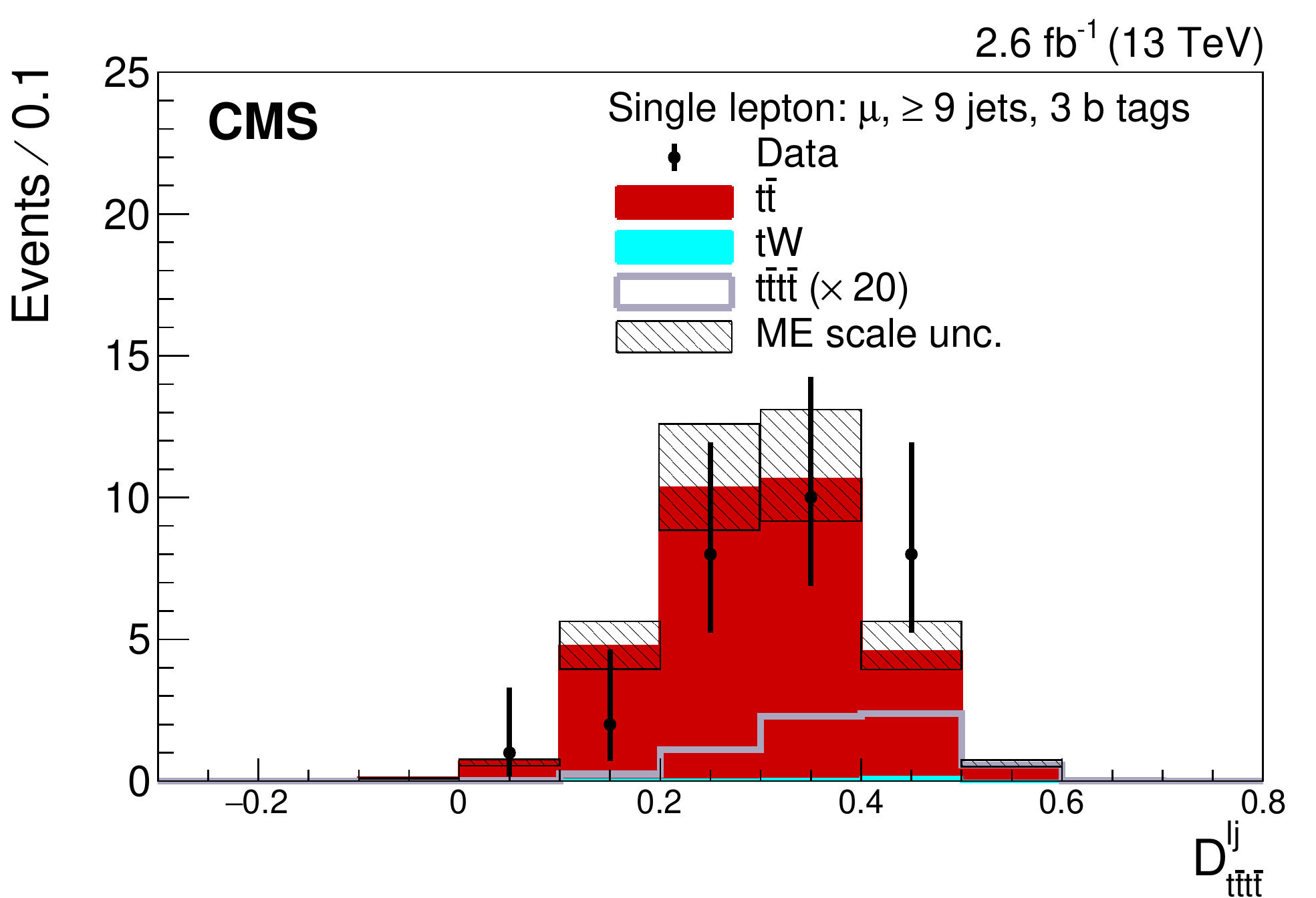}
    \includegraphics[width=0.45\textwidth]{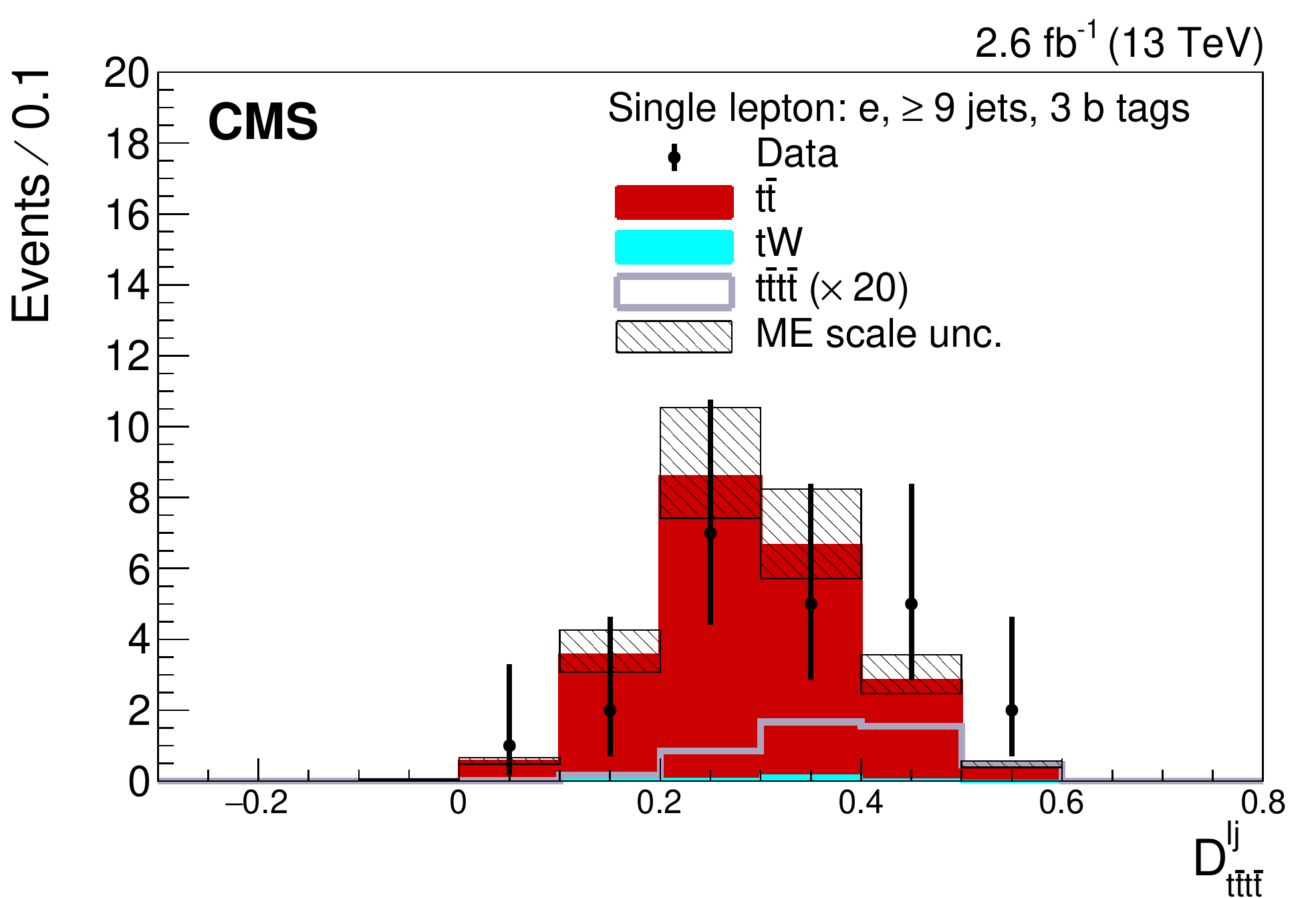}
    \includegraphics[width=0.45\textwidth]{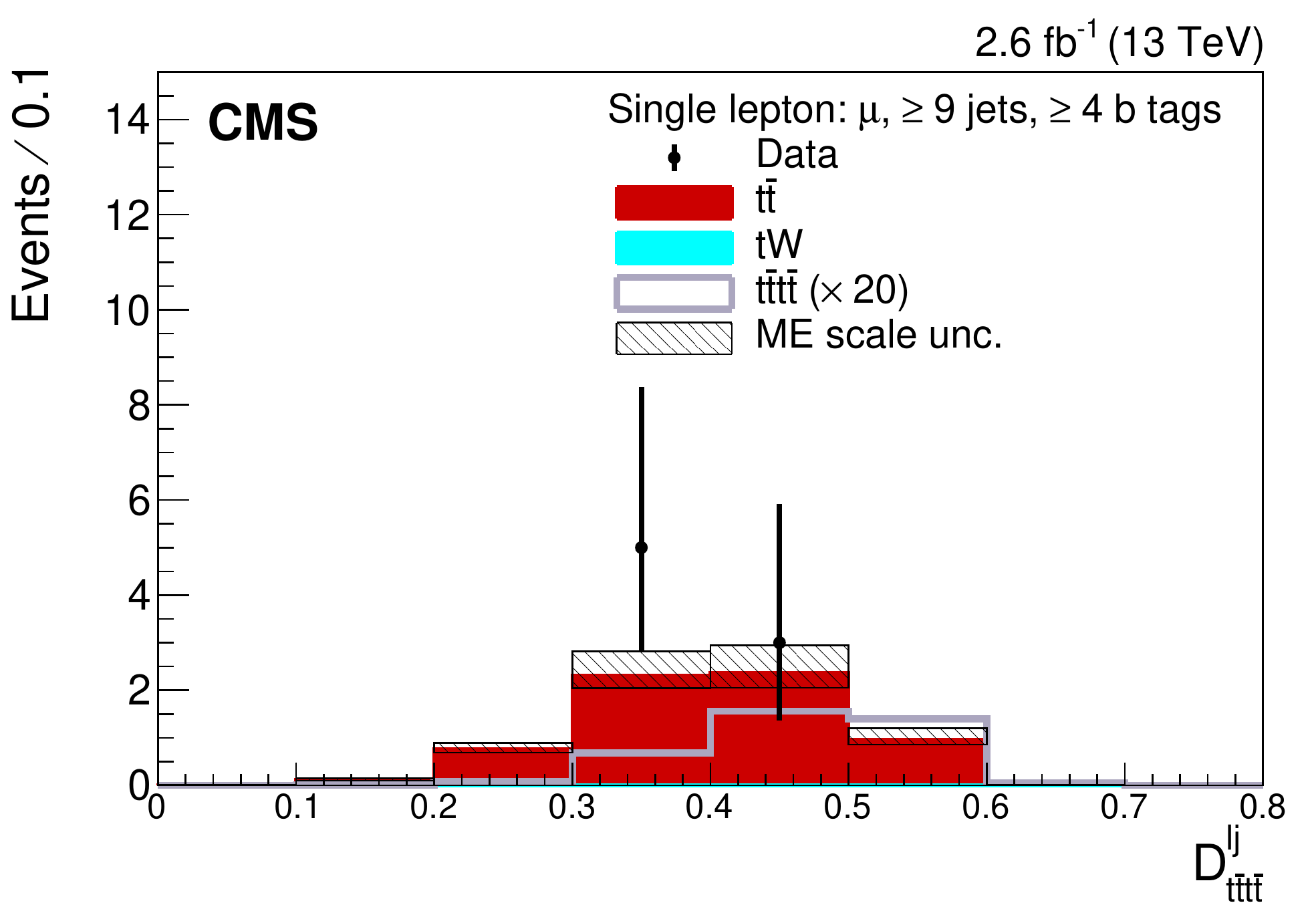}
    \includegraphics[width=0.45\textwidth]{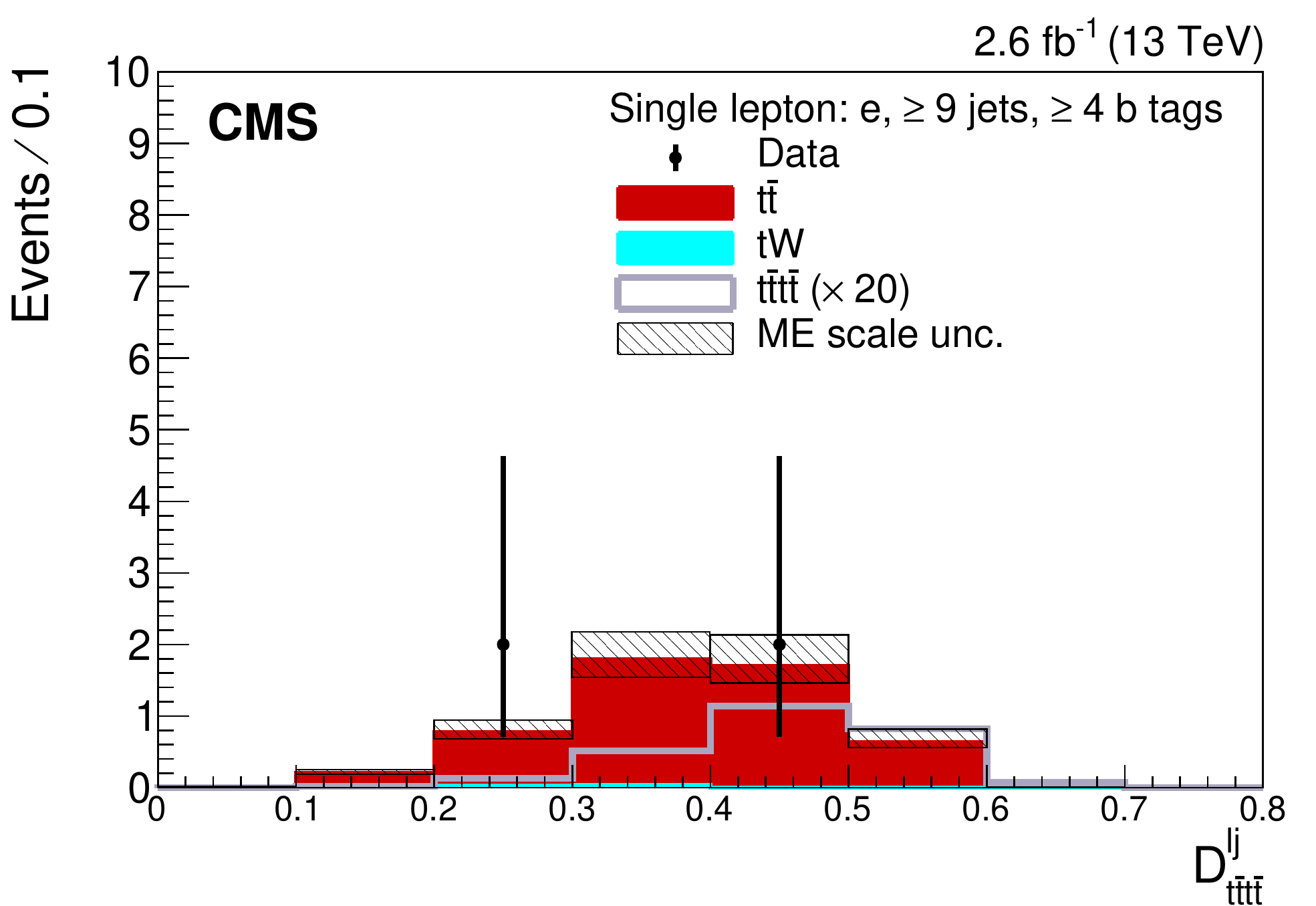}
    \caption{
   	Distribution of the event-level BDT discriminants \BDTljets for the $\mu$+jets (left) and e+jets (right) final states from data and the estimated background contributions from simulation,  in the \njets $\geq9$  and 3 \nmtags (upper panels) and the $\njets \geq 9$ and $\geq$4 $\nmtags$ categories (lower panels). The vertical bars show the statistical uncertainties in the data. The predicted background distributions from simulation are shown by the shaded histograms  The hatched area shows the size of the dominant systematic uncertainty in the simulation, which comes from the matrix-element (ME) factorization and renormalization scales used in the simulation. The expected SM \tttt signal contribution is shown by open histogram, multiplied by a factor of 20.}
    \label{fig:leptBDT}
\end{figure*}

\begin{figure}[!ht]
\centering
\includegraphics[width=0.45\textwidth]{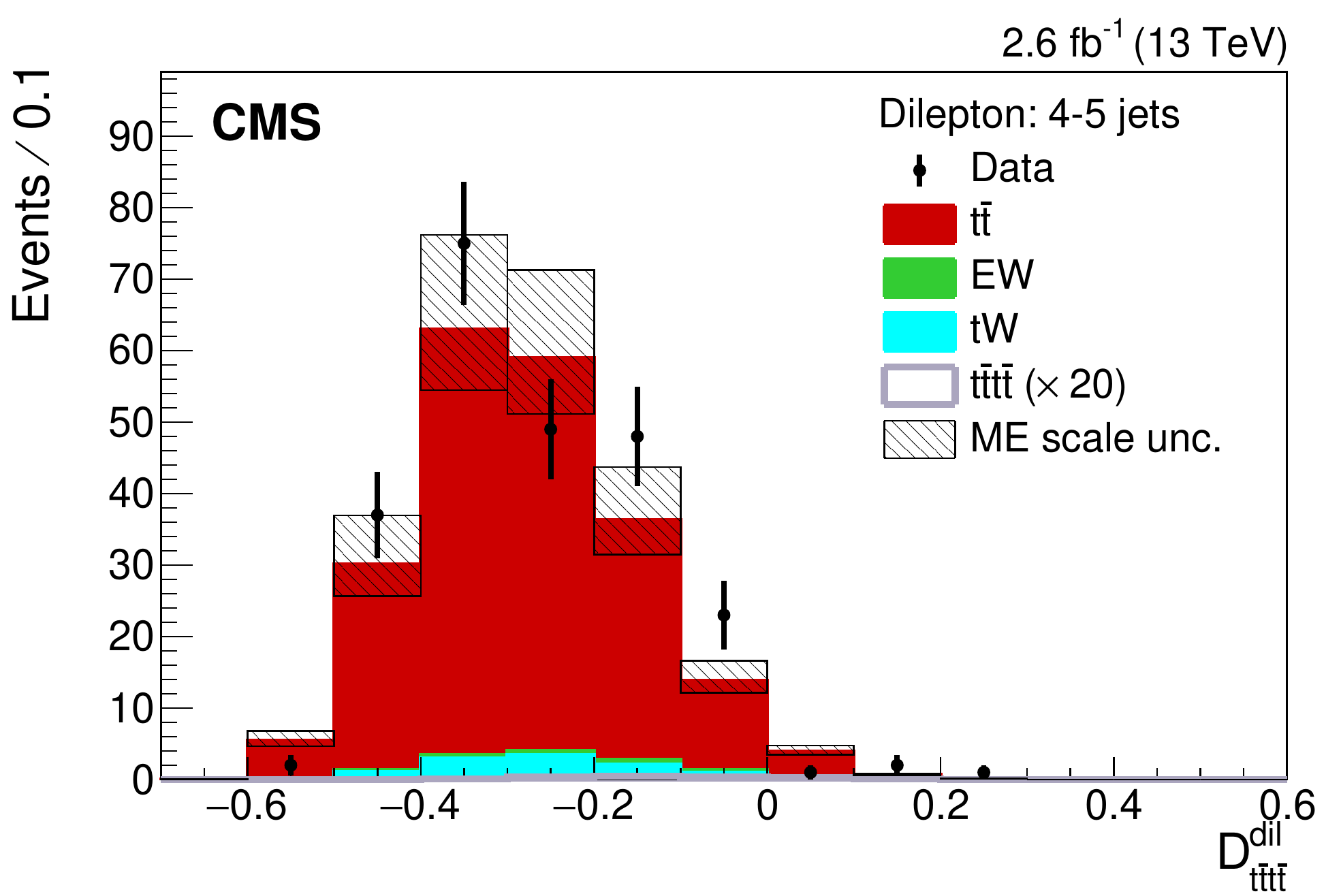}
\includegraphics[width=0.45\textwidth]{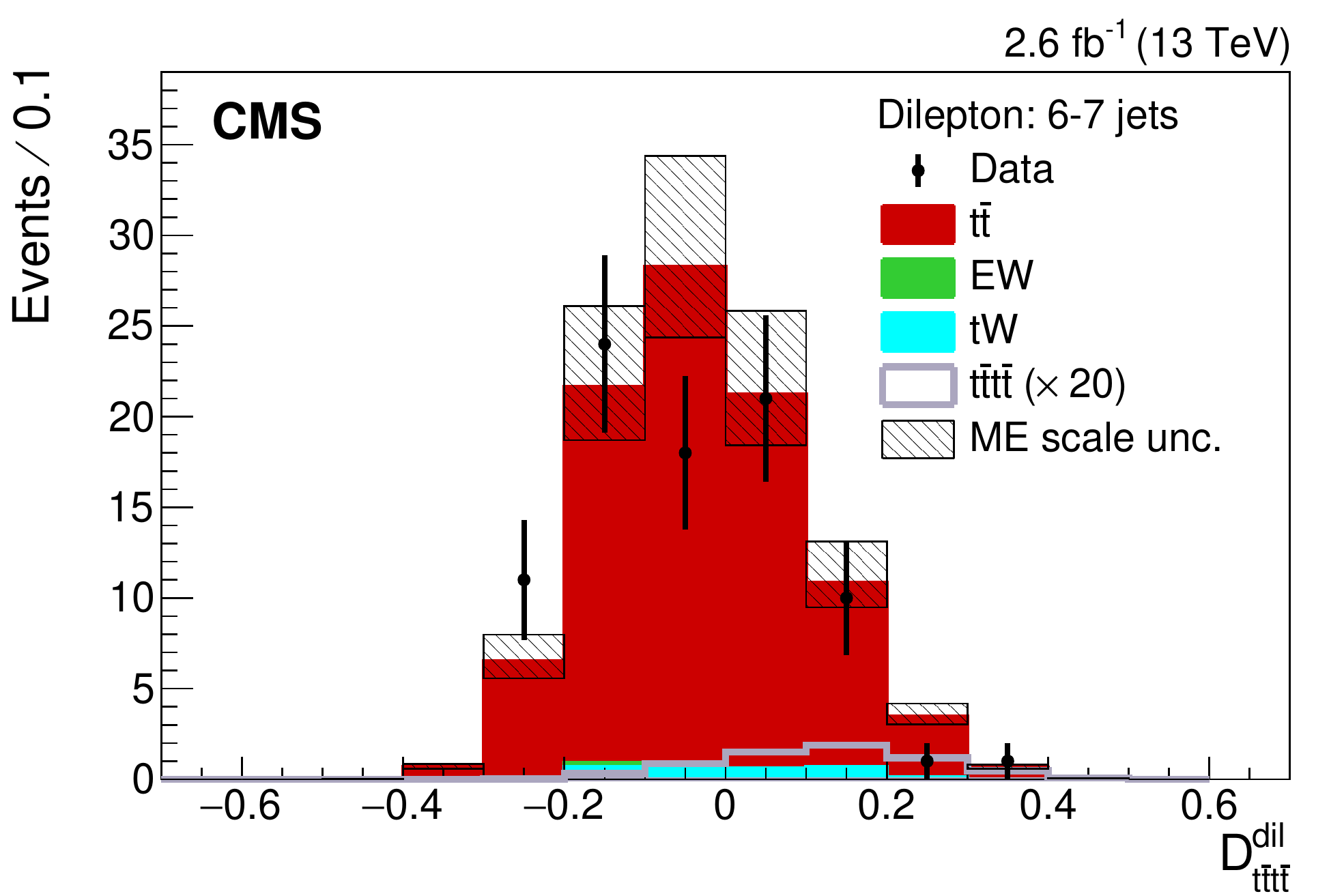}
\includegraphics[width=0.45\textwidth]{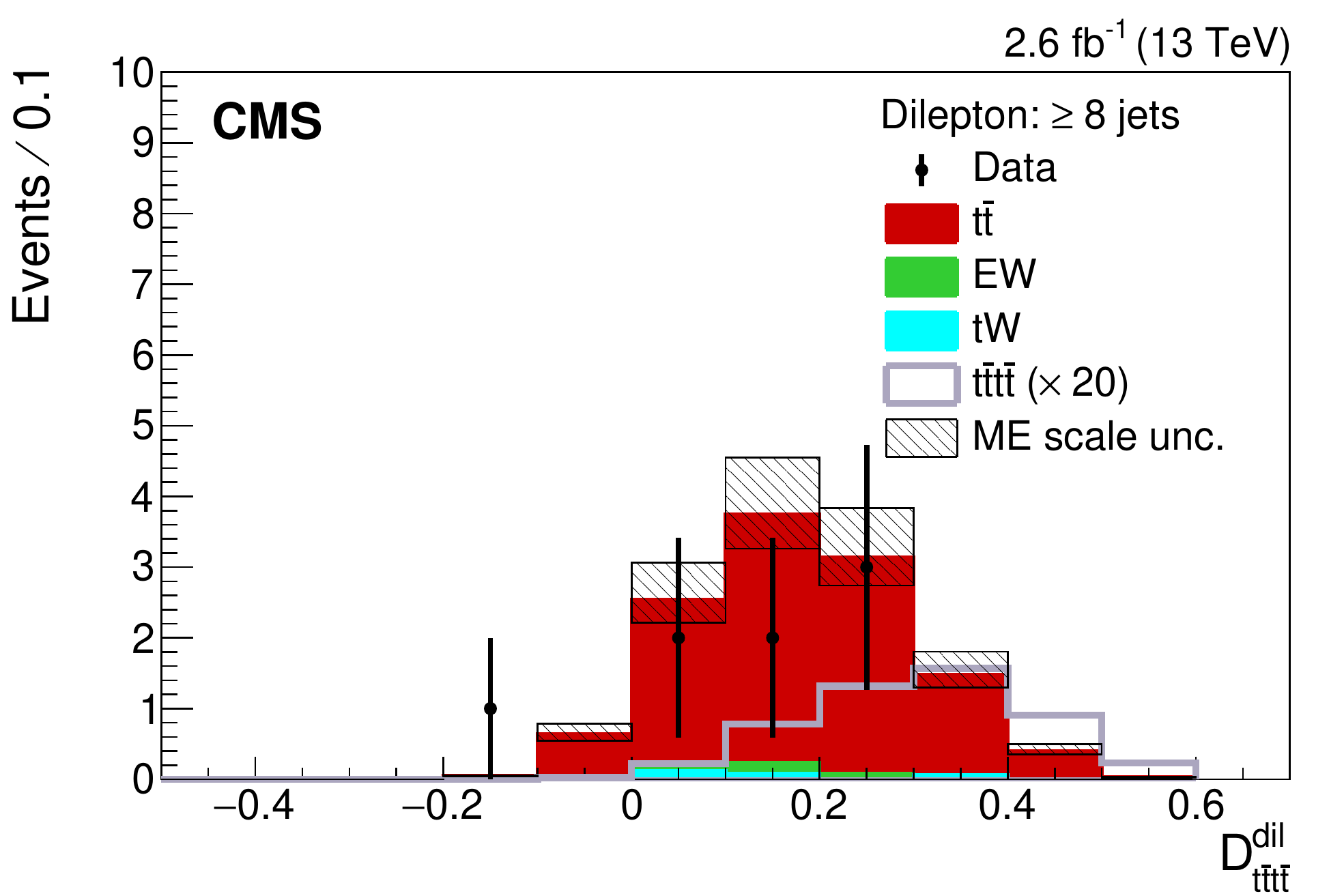}
    \caption{Distribution of the event-level BDT discriminants \BDTdilep for the combined dilepton ($\PGmp\PGmm$+ $\PGm^{\pm} \Pe^{\mp}$+ $\Pep\Pem$) event sample for 4--5 jets (\cmsFirst), 6--7 jets (\cmsMiddle), and $\geq$8 jets (bottom). The vertical bars show the statistical uncertainty in the data. The predicted background distributions from simulation are shown by the shaded histograms. The hatched area shows the size of the dominant systematic uncertainty in the simulation, which comes from the choice of the matrix-element (ME) factorization and renormalization scales used in the simulation. The electroweak (EW) histogram is the sum of the Drell-Yan and W boson+jets backgrounds. The expected SM \tttt signal contribution is shown by the open histogram, multiplied by a factor of 20. }
    \label{fig:dileptBDT}
\end{figure}

To further improve the sensitivity of the analyses, the data are split into exclusive jet multiplicity categories.
The single-lepton analysis uses categories of $\njets=6$, 7, 8, and ${\ge}9$. The dilepton channel,
with fewer events, uses only the 4--5, 6--7, and ${\ge}8$ jet categories.
A further division into exclusive b jet multiplicities is possible only for the single-lepton analysis,
where the \njets categories are subdivided into categories with $\nmtags=2$, 3, and ${\ge}4$.
Figure~\ref{fig:leptBDT} shows \BDTljets in the $\mu$+jets and e+jets channels for two of the most sensitive categories, and Fig.~\ref{fig:dileptBDT} shows the \BDTdilep distributions for the dilepton channel.

The distributions of the discriminants \BDTljets and \BDTdilep are fitted simultaneously for each \njets and \nmtags bin.
For the single-lepton channel the fit is also performed
separately for the  $\PGm$+jets and e+jets events.
In the three dilepton channels ($\PGmp\PGmm$, $\PGm^{\pm} \Pe^{\mp}$, $\Pep\Pem$), the \BDTdilep distributions are found to be consistent, and they are combined to improve the statistical precision. In all cases good agreement is observed between the data and the simulated background, and the results from each of the channels are combined to obtain an upper limit on the \tttt production cross section.

\section{Sources of systematic uncertainty}
The systematic uncertainties affecting the analyses are grouped into normalization and shape categories, depending on their effect on the event-level BDT discriminant distribution.
While all normalization uncertainties apply to both the signal and all the background simulations, the shape uncertainties are only considered for the \ttbar background and the \tttt signal. These include effects related to the change in normalization owing to changes in the shape. The normalization uncertainties are:
\begin{enumerate}
\item An uncertainty of 2.3\%~\cite{CMS-PAS-LUM-15-001} in the integrated luminosity.

\item  The uncertainty in the theoretical \ttbar cross section dominates the uncertainty in the predicted event yields, since the \ttbar process dominates the selected data samples. This cross section is taken from Ref.~\cite{ttNNLO}, and includes uncertainties of $^{+ 2.5\%}_{ - 3.4\%}$ (renormalization and factorization scale) and $^{+ 6.2\%}_{ - 6.4\%}$ (PDF). The effect of uncertainties in the cross sections for the other backgrounds were checked and found to be negligible.

\item The uncertainties from trigger, lepton identification, and lepton isolation corrections, which are included as  nuisance parameters in determining the upper limit. Combined, these give an uncertainty of $1.2\%$ in the single-muon channel, $3.7\%$ in the single-electron channel, $4.3\%$ in the $\mu \mu$ channel, $4.6\%$ in the $\mu $e channel, and $4.8\%$ in the ee channel.
\end{enumerate}

The shape uncertainties are:
\begin{enumerate}
\item The uncertainty from the choice of the factorization and renormalization scales in the calculation of the matrix element of the hard-scattering process, which is estimated by the maximum variation in the \BDTdilep or \BDTljets distribution obtained when each scale is changed separately by a factor of 1/2 and 2, excluding unphysical anticorrelated combinations.
This procedure is performed separately for the \tttt signal and the \ttbar background.
In addition, alternative \ttbar samples are used to estimate the impact of a change in the scale at the parton-shower level, taking into account the uncertainty in $\alpha_{S}$ for the hadronization~\cite{Khachatryan:2015mva}.
The differences in the distributions with respect to the nominal ones are taken as the uncertainty.
The uncertainty in the matrix-element scale is the dominant systematic uncertainty in the analysis.

\item Differences in the simulation of \ttbar from the choice of the matrix-element generator, which is estimated by comparing the nominal \ttbar simulation using \POWHEG{}+\PYTHIA8 to samples generated using \MADGRAPH{}+\PYTHIA8 with MLM matching \cite{Mangano:2006rw}. The difference relative to the nominal simulation is used to estimate the uncertainty from this source.

\item The uncertainty in the fraction of $\PQt\PAQt\PQb\PAQb $ events in the \ttbar background, which is estimated using the uncertainty in the measured cross section ratio
$\sigma_{\PQt\PAQt\PQb\PAQb}$ : $\sigma_{\ttbar\mathrm{jj}}$~\cite{CMS:2016ttbb} that was used to correct the
$\PQt\PAQt\PQb\PAQb$ content of the \ttbar simulation.
An anticorrelated uncertainty in the measured cross section ratio of
$(\sigma_{\ttbar\mathrm{jj}} - \sigma_{\PQt\PAQt\PQb\PAQb}$) : $\sigma_{\ttbar\mathrm{jj}}$ is applied simultaneously to the light-quark fraction to maintain the total \ttbar cross section.

\item The uncertainties in the jet energy scale and the jet energy resolution \cite{Khachatryan:2016kdb}, which are estimated by varying these within their uncertainties by $\pm$1 standard deviation.
 A similar method is used to estimate the uncertainty from the inelastic proton-proton cross section and the procedure used in the pileup reweighting. These uncertainties have very little influence on the final limit.

\item The uncertainty in the corrections to the values of the b tagging CSV discriminator, where three categories of systematic uncertainty are applied for each jet flavor: the jet energy scale, purity of the data sample used to derive the corrections, and the statistical uncertainties derived from the fits used in the method.
	The uncertainty in the b tagging correction caused by the jet energy scale is treated as fully correlated with the jet energy scale uncertainty described above.
Typical magnitudes of each of these individual uncertainties on the corrections to the b tagging CSV discriminator are 10--50\% before the fit, depending on the number of jets and b jets in the event.
A full description of these corrections can be found in Ref.~\cite{Khachatryan:2014qaa}.
\end{enumerate}
Each systematic source was attributed a nuisance parameter in the limit determination.

\section{Results}
No deviation from the background-only simulation, which includes \ttbar production and negligible single top, \ttbar{}+H/Z/W boson, Drell--Yan+jets, and W+jets backgrounds, is observed in the \BDTdilep or \BDTljets distributions.
An upper limit is derived for the \tttt production cross section using the asymptotic approximation of the $CL_{\rm{s}}$ method provided in Refs.~\cite{RooStats:2010,Read:2002hq,Junk:1999kv,Cowan:2011js, CMS-NOTE-2011-005}.
The signal and background distributions are fitted using a simultaneous maximum-likelihood method.
The normalization uncertainties are included using log-normal functions and the shape uncertainties are included as Gaussian-distributed nuisance parameters.
The expected and observed 95\% CL upper limits from the two analyses and their combination are listed in Table~\ref{tab:limits_single}.
For the combination of the single-lepton and opposite-sign dilepton results, the systematic uncertainties attributed to the integrated luminosity, jet energy scale, and modeling of the pileup contribution are assumed to be fully correlated. All other systematic uncertainties are assumed to be uncorrelated, but taking them as fully correlated does not modify the expected limit. The likelihood function for the single and opposite-sign dilepton limit has 24 nuisance parameters corresponding to the sources of systematic uncertainties that are described above. The combination with the like-sign dilepton analysis, which is described below, adds 19 additional nuisance parameters specific to Ref.~\cite{Khachatryan:2016kod}. The data are able to significantly constrain the parameters corresponding to the ME generator choice and the parton shower scale. All of the post-fit nuisance parameter values were found to be consistent with their initial values to well within their quoted uncertainties.

\begin{table*}[ht!]
	\topcaption{Expected and observed 95\% CL upper limits on SM \tttt production as a multiple of \sigmattttSM and in fb. The results for the two analyses from this paper are shown separately and combined. The result from a previous CMS measurement  \cite{Khachatryan:2016kod} is also given, along with the overall limits when the three measurements are combined. The values quoted for the uncertainties on the expected limits are the one standard deviation values and include all statistical and systematic uncertainties.}	
	\centering
	\begin{tabular}{ l  c   c  c  c }
		Channel  & Expected limit  & Observed limit & Expected limit  & Observed limit \T \\	
		 & ($\times$\sigmattttSM) & ($\times$\sigmattttSM) & (fb) & (fb) \T \B \\ \hline
                Single lepton  & $\xsecmusingleptonexp^{\,+\,\xsecmusingleptonup}_{\,-\,\xsecmusingleptondown}$ & $\xsecmusinglepton$ & $\xsecfbsingleptonexp^{\,+\,\xsecfbsingleptonup}_{\,-\,\xsecfbsingleptondown}$ & $\xsecfbsinglepton$   \T \B  \\
                Dilepton  & $\xsecmudileptonexp^{\,+\,\xsecmudileptonup}_{\,-\,\xsecmudileptondown}$ & $\xsecmudilepton$ & $\xsecfbdileptonexp^{\,+\,\xsecfbdileptonup}_{\,-\,\xsecfbdileptondown}$ & $\xsecfbdilepton$ \T   \\
                (opposite sign) & & & &  \\[1.5ex]
                 Combined  & $\xsecmucomboexp^{\,+\,\xsecmucomboup}_{\,-\,\xsecmucombodown}$ & $\xsecmucombo$  & $\xsecfbcomboexp^{\,+\,\xsecfbcomboup}_{\,-\,\xsecfbcombodown}$ & $\xsecfbcombo$   \T   \\
                (this analysis) & & & &  \\   \hline
                Dilepton & $11.0^{\,+\,6.2}_{\,-\,3.8}$ & $12.9$ & $101^{\,+\,57}_{\,-\,35}$ & $119$   \T   \\
                (same sign  \cite{Khachatryan:2016kod} ) & & &  & \\[1.5ex]
                Combined  & $\xsecmucomboallexp^{\,+\,\xsecmucomboallup}_{\,-\,\xsecmucomboalldown}$ & $\xsecmucomboall$  & $\xsecfbcomboallexp^{\,+\,\xsecfbcomboallup}_{\,-\,\xsecfbcomboalldown}$ & $\xsecfbcomboall$  \T \B   \\
	\end{tabular}
	\label{tab:limits_single}
\end{table*}

The combined observed 95\% CL upper limits on the cross section for four-top-quark production measured in the single-lepton, dilepton, and combined results are shown in Table~\ref{tab:limits_single}.
To cross-check the increase in sensitivity of the multivariate approach, the analysis is performed in the single-lepton channel using the same event categorization, but only the \HT distributions in place of \BDTljets.  The expected limit increases by approximately 20\%, thus justifying the use of the more complicated BDT analyses.

CMS has also produced an upper limit on the  \tttt cross section from the analysis of the like-sign dilepton channel~\cite{Khachatryan:2016kod}. The limit from the analysis is also shown in Table~\ref{tab:limits_single}.
To improve sensitivity, the results from this search are combined with the results from that analysis. For this combination, in addition to the assumed correlations described above, the uncertainty in the modeling of the response of the CMS trigger system to dilepton events is treated as correlated between the opposite-sign and like-sign dilepton analyses.
A combined upper limit for the SM \tttt cross section is listed in Table~\ref{tab:limits_single}.

\section{Summary}

In summary, a search has been performed for events containing four top quarks using data recorded by the CMS experiment in proton-proton collisions at $\sqrt{s} = 13$\TeV corresponding to an integrated luminosity of \ourLumi.
The final states considered in the analysis are the single-lepton channel with exactly one electron or muon, and the opposite-sign dilepton channel with exactly two of any combination of electrons or muons.
A boosted decision tree is used to discriminate between the \tttt signal and the \ttbar background, and no signal is observed.
This leads to an upper limit on the SM production cross section for \tttt of $\xsecfbcombo\unit{fb}$ ($\xsecmucombo \, \sigmattttSM$), with an expected limit of $\xsecfbcomboexp^{\,+\,\xsecfbcomboup}_{\,-\,\xsecfbcombodown}\unit{fb}$ at the 95\% confidence level.
This result is combined with a previous search \cite{Khachatryan:2016kod} with similar sensitivity in the same-sign dilepton channel to obtain an improved limit of  $\xsecfbcomboall$\unit{fb},
with an expected limit of $\xsecfbcomboallexp^{\,+\,\xsecfbcomboallup}_{\,-\,\xsecfbcomboalldown}$\unit{fb}.
This is the most stringent limit on \tttt production at  $\sqrt{s} = 13$\TeV published to date.

\begin{acknowledgments}
We congratulate our colleagues in the CERN accelerator departments for the excellent performance of the LHC and thank the technical and administrative staffs at CERN and at other CMS institutes for their contributions to the success of the CMS effort. In addition, we gratefully acknowledge the computing centers and personnel of the Worldwide LHC Computing Grid for delivering so effectively the computing infrastructure essential to our analyses. Finally, we acknowledge the enduring support for the construction and operation of the LHC and the CMS detector provided by the following funding agencies: BMWFW and FWF (Austria); FNRS and FWO (Belgium); CNPq, CAPES, FAPERJ, and FAPESP (Brazil); MES (Bulgaria); CERN; CAS, MoST, and NSFC (China); COLCIENCIAS (Colombia); MSES and CSF (Croatia); RPF (Cyprus); SENESCYT (Ecuador); MoER, ERC IUT, and ERDF (Estonia); Academy of Finland, MEC, and HIP (Finland); CEA and CNRS/IN2P3 (France); BMBF, DFG, and HGF (Germany); GSRT (Greece); OTKA and NIH (Hungary); DAE and DST (India); IPM (Iran); SFI (Ireland); INFN (Italy); MSIP and NRF (Republic of Korea); LAS (Lithuania); MOE and UM (Malaysia); BUAP, CINVESTAV, CONACYT, LNS, SEP, and UASLP-FAI (Mexico); MBIE (New Zealand); PAEC (Pakistan); MSHE and NSC (Poland); FCT (Portugal); JINR (Dubna); MON, RosAtom, RAS, RFBR and RAEP (Russia); MESTD (Serbia); SEIDI, CPAN, PCTI and FEDER (Spain); Swiss Funding Agencies (Switzerland); MST (Taipei); ThEPCenter, IPST, STAR, and NSTDA (Thailand); TUBITAK and TAEK (Turkey); NASU and SFFR (Ukraine); STFC (United Kingdom); DOE and NSF (USA).

\hyphenation{Rachada-pisek} Individuals have received support from the Marie-Curie program and the European Research Council and EPLANET (European Union); the Leventis Foundation; the A. P. Sloan Foundation; the Alexander von Humboldt Foundation; the Belgian Federal Science Policy Office; the Fonds pour la Formation \`a la Recherche dans l'Industrie et dans l'Agriculture (FRIA-Belgium); the Agentschap voor Innovatie door Wetenschap en Technologie (IWT-Belgium); the Ministry of Education, Youth and Sports (MEYS) of the Czech Republic; the Council of Science and Industrial Research, India; the HOMING PLUS program of the Foundation for Polish Science, cofinanced from European Union, Regional Development Fund, the Mobility Plus program of the Ministry of Science and Higher Education, the National Science Center (Poland), contracts Harmonia 2014/14/M/ST2/00428, Opus 2014/13/B/ST2/02543, 2014/15/B/ST2/03998, and 2015/19/B/ST2/02861, Sonata-bis 2012/07/E/ST2/01406; the National Priorities Research Program by Qatar National Research Fund; the Programa Clar\'in-COFUND del Principado de Asturias; the Thalis and Aristeia programs cofinanced by EU-ESF and the Greek NSRF; the Rachadapisek Sompot Fund for Postdoctoral Fellowship, Chulalongkorn University and the Chulalongkorn Academic into Its 2nd Century Project Advancement Project (Thailand); and the Welch Foundation, contract C-1845. \end{acknowledgments}
\bibliography{auto_generated}

\providecommand{\href}[2]{#2}\begingroup\raggedright\begin{thebibliography}{10}%
\makeatletter
\providecommand{\hrefCMSnoop }[0]{\@secondoftwo}%
\makeatother
\providecommand{\doi}{\texttt{doi:}\begingroup \urlstyle{tt}\Url}

\bibitem{hdmrev}
\hrefCMSnoop {}{G.~C. Branco {et~al.}, ``Theory and phenomenology of
  two-{H}iggs-doublet models'',} \textit{ Phys. Rep. D} \textbf{ 516} (2012) 1,
  \href{http://dx.doi.org/10.1016/j.physrep.2012.02.002}{\doi{10.1016/j.physrep.2012.02.002}},
  \href{http://www.arXiv.org/abs/1106.0034}{\texttt{arXiv:1106.0034}}.

\bibitem{Toharia:2005gm}
\hrefCMSnoop {}{M.~Toharia and J.~D. Wells, ``{Gluino decays with heavier
  scalar superpartners}'',} \textit{ JHEP} \textbf{ 02} (2006) 015,
  \href{http://dx.doi.org/10.1088/1126-6708/2006/02/015}{\doi{10.1088/1126-6708/2006/02/015}},
\href{http://www.arXiv.org/abs/hep-ph/0503175}{\texttt{arXiv:hep-ph/0503175}}.

\bibitem{Kumar:2009vs}
\hrefCMSnoop {}{K.~Kumar, T.~M.~P. Tait, and R.~Vega-Morales, ``{Manifestations
  of top compositeness at colliders}'',} \textit{ JHEP} \textbf{ 05} (2009)
  022,
  \href{http://dx.doi.org/10.1088/1126-6708/2009/05/022}{\doi{10.1088/1126-6708/2009/05/022}},
\href{http://www.arXiv.org/abs/0901.3808}{\texttt{arXiv:0901.3808}}.

\bibitem{Calvet:2012rk}
\hrefCMSnoop {}{S.~Calvet, B.~Fuks, P.~Gris, and L.~Valery, ``{Searching for
  sgluons in multitop events at a center-of-mass energy of 8 TeV}'',} \textit{
  JHEP} \textbf{ 04} (2013) 043,
  \href{http://dx.doi.org/10.1007/JHEP04(2013)043}{\doi{10.1007/JHEP04(2013)043}},
\href{http://www.arXiv.org/abs/1212.3360}{\texttt{arXiv:1212.3360}}.

\bibitem{Cacciapaglia:2015eqa}
G.~Cacciapaglia\hrefCMSnoop {}{ {et~al.}, ``{Composite scalars at the LHC: the
  Higgs, the Sextet and the Octet}'',} \textit{ JHEP} \textbf{ 11} (2015) 201,
  \href{http://dx.doi.org/10.1007/JHEP11(2015)201}{\doi{10.1007/JHEP11(2015)201}},
\href{http://www.arXiv.org/abs/1507.02283}{\texttt{arXiv:1507.02283}}.

\bibitem{Ducu:2015fda}
\hrefCMSnoop {}{O.~Ducu, L.~Heurtier, and J.~Maurer, ``{LHC signatures of a
  $\mathrm{Z}'$ mediator between dark matter and the SU(3) sector}'',} \textit{
  JHEP} \textbf{ 03} (2016) 006,
  \href{http://dx.doi.org/10.1007/JHEP03(2016)006}{\doi{10.1007/JHEP03(2016)006}},
\href{http://www.arXiv.org/abs/1509.05615}{\texttt{arXiv:1509.05615}}.

\bibitem{Arina:2016cqj}
C.~Arina\hrefCMSnoop {}{ {et~al.}, ``{A comprehensive approach to dark matter
  studies: exploration of simplified top-philic models}'',} \textit{ JHEP}
  \textbf{ 11} (2016) 111,
  \href{http://dx.doi.org/10.1007/JHEP11(2016)111}{\doi{10.1007/JHEP11(2016)111}},
\href{http://www.arXiv.org/abs/1605.09242}{\texttt{arXiv:1605.09242}}.

\bibitem{Cao:2016wib}
\hrefCMSnoop {}{Q.-H. Cao, S.-L. Chen, and Y.~Liu, ``{Probing Higgs width and
  top quark Yukawa coupling from $\textrm{t}\bar{\textrm{t}}\textrm{H}$ and
  $\textrm{t}\bar{\textrm{t}}\textrm{t}\bar{\textrm{t}}$ productions}'',}
  \textit{ Phys. Rev. D} \textbf{ 95} (2017) 053004,
  \href{http://dx.doi.org/10.1103/PhysRevD.95.053004}{\doi{10.1103/PhysRevD.95.053004}},
\href{http://www.arXiv.org/abs/1602.01934}{\texttt{arXiv:1602.01934}}.

\bibitem{Bevilacqua:2012em}
\hrefCMSnoop {}{G.~Bevilacqua and M.~Worek, ``{Constraining BSM physics at the
  LHC: Four top final states with NLO accuracy in perturbative QCD}'',}
  \textit{ JHEP} \textbf{ 07} (2012) 111,
  \href{http://dx.doi.org/10.1007/JHEP07(2012)111}{\doi{10.1007/JHEP07(2012)111}},
\href{http://www.arXiv.org/abs/1206.3064}{\texttt{arXiv:1206.3064}}.

\bibitem{Khachatryan2014}
\hrefCMSnoop {}{{CMS Collaboration}, ``Search for standard model production of
  four top quarks in the lepton+jets channel in pp collisions at {$\sqrt{s} =
  8\TeV$}'',} \textit{ JHEP} \textbf{ 11} (2014) 154,
  \href{http://dx.doi.org/10.1007/JHEP11(2014)154}{\doi{10.1007/JHEP11(2014)154}},
  \href{http://www.arXiv.org/abs/1409.7339}{\texttt{arXiv:1409.7339}}.

\bibitem{Aad:2015kqa}
\hrefCMSnoop {}{{ATLAS Collaboration}, ``{Search for production of vector-like
  quark pairs and of four top quarks in the lepton-plus-jets final state in pp
  collisions at $\sqrt{s}=8$ TeV with the ATLAS detector}'',} \textit{ JHEP}
  \textbf{ 08} (2015) 105,
  \href{http://dx.doi.org/10.1007/JHEP08(2015)105}{\doi{10.1007/JHEP08(2015)105}},
\href{http://www.arXiv.org/abs/1505.04306}{\texttt{arXiv:1505.04306}}.

\bibitem{Alwall:2014hca}
J.~Alwall\hrefCMSnoop {}{ {et~al.}, ``{The automated computation of tree-level
  and next-to-leading order differential cross sections, and their matching to
  parton shower simulations}'',} \textit{ JHEP} \textbf{ 07} (2014) 079,
  \href{http://dx.doi.org/10.1007/JHEP07(2014)079}{\doi{10.1007/JHEP07(2014)079}},
\href{http://www.arXiv.org/abs/1405.0301}{\texttt{arXiv:1405.0301}}.

\bibitem{Khachatryan:2016kod}
\hrefCMSnoop {}{{CMS Collaboration}, ``{Search for new physics in same-sign
  dilepton events in proton-proton collisions at $\sqrt{s}$ = 13 TeV}'',}
  \textit{ Eur. Phys. J. C} \textbf{ 76} (2016) 439,
  \href{http://dx.doi.org/10.1140/epjc/s10052-016-4261-z}{\doi{10.1140/epjc/s10052-016-4261-z}},
\href{http://www.arXiv.org/abs/1605.03171}{\texttt{arXiv:1605.03171}}.

\bibitem{Khachatryan:2016bia}
\hrefCMSnoop {}{{CMS Collaboration}, ``{The CMS trigger system}'',} \textit{
  JINST} \textbf{ 12} (2017) P01020,
  \href{http://dx.doi.org/10.1088/1748-0221/12/01/P01020}{\doi{10.1088/1748-0221/12/01/P01020}},
\href{http://www.arXiv.org/abs/1609.02366}{\texttt{arXiv:1609.02366}}.

\bibitem{CMSdet}
\hrefCMSnoop {}{{CMS Collaboration}, ``The {CMS} experiment at the {CERN}
  {LHC}'',} \textit{ JINST} \textbf{ 3} (2008) S08004,
\href{http://dx.doi.org/10.1088/1748-0221/3/08/S08004}{\doi{10.1088/1748-0221/3/08/S08004}}.

\bibitem{Mangano:2006rw}
\hrefCMSnoop {}{M.~L. Mangano, M.~Moretti, F.~Piccinini, and M.~Treccani,
  ``{Matching matrix elements and shower evolution for top-quark production in
  hadronic collisions}'',} \textit{ JHEP} \textbf{ 01} (2007) 013,
  \href{http://dx.doi.org/10.1088/1126-6708/2007/01/013}{\doi{10.1088/1126-6708/2007/01/013}},
\href{http://www.arXiv.org/abs/hep-ph/0611129}{\texttt{arXiv:hep-ph/0611129}}.

\bibitem{Nason:2004rx}
\hrefCMSnoop {}{P.~Nason, ``{A new method for combining NLO QCD with shower
  Monte Carlo algorithms}'',} \textit{ JHEP} \textbf{ 11} (2004) 040,
  \href{http://dx.doi.org/10.1088/1126-6708/2004/11/040}{\doi{10.1088/1126-6708/2004/11/040}},
\href{http://www.arXiv.org/abs/hep-ph/0409146}{\texttt{arXiv:hep-ph/0409146}}.

\bibitem{Frixione:2007vw}
\hrefCMSnoop {}{S.~Frixione, P.~Nason, and C.~Oleari, ``{Matching NLO QCD
  computations with parton shower simulations: the POWHEG method}'',} \textit{
  JHEP} \textbf{ 11} (2007) 070,
  \href{http://dx.doi.org/10.1088/1126-6708/2007/11/070}{\doi{10.1088/1126-6708/2007/11/070}},
\href{http://www.arXiv.org/abs/0709.2092}{\texttt{arXiv:0709.2092}}.

\bibitem{Alioli:2010xd}
\hrefCMSnoop {}{S.~Alioli, P.~Nason, C.~Oleari, and E.~Re, ``{A general
  framework for implementing NLO calculations in shower Monte Carlo programs:
  the POWHEG BOX}'',} \textit{ JHEP} \textbf{ 06} (2010) 043,
  \href{http://dx.doi.org/10.1007/JHEP06(2010)043}{\doi{10.1007/JHEP06(2010)043}},
\href{http://www.arXiv.org/abs/1002.2581}{\texttt{arXiv:1002.2581}}.

\bibitem{Alioli2012}
\hrefCMSnoop {}{S.~Alioli, S.-O. Moch, and P.~Uwer, ``Hadronic top-quark
  pair-production with one jet and parton showering'',} \textit{ JHEP} \textbf{
  01} (2012) 137,
  \href{http://dx.doi.org/10.1007/JHEP01(2012)137}{\doi{10.1007/JHEP01(2012)137}},
  \href{http://www.arXiv.org/abs/1110.5251}{\texttt{arXiv:1110.5251}}.

\bibitem{Alwall:2008}
J.~Alwall\hrefCMSnoop {}{ {et~al.}, ``Comparative study of various algorithms
  for the merging of parton showers and matrix elements in hadronic
  collisions'',} \textit{ Eur. Phys. J. C} \textbf{ 53} (2008) 473,
  \href{http://dx.doi.org/10.1140/epjc/s10052-007-0490-5}{\doi{10.1140/epjc/s10052-007-0490-5}},
  \href{http://www.arXiv.org/abs/0706.2569}{\texttt{arXiv:0706.2569}}.

\bibitem{1126-6708-2006-05-026}
\hrefCMSnoop {}{T.~Sj{\"o}strand, S.~Mrenna, and P.~Skands, ``{PYTHIA} 6.4
  physics and manual'',} \textit{ JHEP} \textbf{ 05} (2006) 026,
  \href{http://dx.doi.org/10.1088/1126-6708/2006/05/026}{\doi{10.1088/1126-6708/2006/05/026}},
\href{http://www.arXiv.org/abs/hep-ph/0603175}{\texttt{arXiv:hep-ph/0603175}}.

\bibitem{Sjostrand2008852}
\hrefCMSnoop {}{T.~Sj{\"o}strand, S.~Mrenna, and P.~Skands, ``A brief
  introduction to {PYTHIA} 8.1'',} \textit{ Comp. Phys. Comm.} \textbf{ 178}
  (2008) 852,
  \href{http://dx.doi.org/10.1016/j.cpc.2008.01.036}{\doi{10.1016/j.cpc.2008.01.036}},
\href{http://www.arXiv.org/abs/0710.3820}{\texttt{arXiv:0710.3820}}.

\bibitem{Khachatryan:2015pea}
\hrefCMSnoop {}{{CMS Collaboration}, ``{Event generator tunes obtained from
  underlying event and multiparton scattering measurements}'',} \textit{ Eur.
  Phys. J. C} \textbf{ 76} (2016) 155,
  \href{http://dx.doi.org/10.1140/epjc/s10052-016-3988-x}{\doi{10.1140/epjc/s10052-016-3988-x}},
\href{http://www.arXiv.org/abs/1512.00815}{\texttt{arXiv:1512.00815}}.

\bibitem{Khachatryan:2015mva}
\hrefCMSnoop {}{{CMS Collaboration}, ``Measurement of $\mathrm
  {t}\overline{\mathrm {t}}$ production with additional jet activity, including
  $\mathrm {b}$ quark jets, in the dilepton decay channel using pp collisions
  at {$\sqrt{s} = 8$\TeV}'',} \textit{ Eur. Phys. J. C} \textbf{ 76} (2016)
  379,
  \href{http://dx.doi.org/10.1140/epjc/s10052-016-4105-x}{\doi{10.1140/epjc/s10052-016-4105-x}},
\href{http://www.arXiv.org/abs/1510.03072}{\texttt{arXiv:1510.03072}}.

\bibitem{Ball:2011uy}
\hrefCMSnoop {}{{NNPDF} Collaboration, ``{Unbiased global determination of
  parton distributions and their uncertainties at NNLO and at LO}'',} \textit{
  Nucl. Phys. B} \textbf{ 855} (2012) 153,
  \href{http://dx.doi.org/10.1016/j.nuclphysb.2011.09.024}{\doi{10.1016/j.nuclphysb.2011.09.024}},
\href{http://www.arXiv.org/abs/1107.2652}{\texttt{arXiv:1107.2652}}.

\bibitem{Agostinelli2003250}
\hrefCMSnoop {}{{GEANT4} Collaboration, ``{GEANT4}---a simulation toolkit'',}
  \textit{ Nucl. Instrum. Meth. A} \textbf{ 506} (2003) 250,
\href{http://dx.doi.org/10.1016/S0168-9002(03)01368-8}{\doi{10.1016/S0168-9002(03)01368-8}}.

\bibitem{Czakon:2011xx}
\hrefCMSnoop {}{M.~Czakon and A.~Mitov, ``{Top++: A program for the calculation
  of the top-pair cross-section at hadron colliders}'',} \textit{ Comput. Phys.
  Commun.} \textbf{ 185} (2014) 2930,
  \href{http://dx.doi.org/10.1016/j.cpc.2014.06.021}{\doi{10.1016/j.cpc.2014.06.021}},
\href{http://www.arXiv.org/abs/1112.5675}{\texttt{arXiv:1112.5675}}.

\bibitem{Aliev:2010zk}
M.~Aliev\hrefCMSnoop {}{ {et~al.}, ``{HATHOR: HAdronic Top and Heavy quarks
  crOss section calculatoR}'',} \textit{ Comput. Phys. Commun.} \textbf{ 182}
  (2011) 1034,
  \href{http://dx.doi.org/10.1016/j.cpc.2010.12.040}{\doi{10.1016/j.cpc.2010.12.040}},
\href{http://www.arXiv.org/abs/1007.1327}{\texttt{arXiv:1007.1327}}.

\bibitem{Kant:2014oha}
P.~Kant\hrefCMSnoop {}{ {et~al.}, ``{HatHor for single top-quark production:
  Updated predictions and uncertainty estimates for single top-quark production
  in hadronic collisions}'',} \textit{ Comput. Phys. Commun.} \textbf{ 191}
  (2015) 74,
  \href{http://dx.doi.org/10.1016/j.cpc.2015.02.001}{\doi{10.1016/j.cpc.2015.02.001}},
\href{http://www.arXiv.org/abs/1406.4403}{\texttt{arXiv:1406.4403}}.

\bibitem{Campbell:2006wx}
\hrefCMSnoop {}{J.~M. Campbell, J.~W. Huston, and W.~J. Stirling, ``{Hard
  interactions of quarks and gluons: A primer for LHC physics}'',} \textit{
  Rept. Prog. Phys.} \textbf{ 70} (2007) 89,
  \href{http://dx.doi.org/10.1088/0034-4885/70/1/R02}{\doi{10.1088/0034-4885/70/1/R02}},
\href{http://www.arXiv.org/abs/hep-ph/0611148}{\texttt{arXiv:hep-ph/0611148}}.

\bibitem{Khachatryan:2015hwa}
\hrefCMSnoop {}{{CMS Collaboration}, ``Performance of electron reconstruction
  and selection with the {CMS} detector in proton-proton collisions at
  {$\sqrt{s} = 8$\TeV}'',} \textit{ JINST} \textbf{ 10} (2015) P06005,
  \href{http://dx.doi.org/10.1088/1748-0221/10/06/P06005}{\doi{10.1088/1748-0221/10/06/P06005}},
  \href{http://www.arXiv.org/abs/1502.02701}{\texttt{arXiv:1502.02701}}.

\bibitem{CMS-PAPER-MUO-10-004}
\hrefCMSnoop {}{{CMS Collaboration}, ``Performance of {CMS} muon reconstruction
  in pp collision events at {$\sqrt{s} = 7$\TeV}'',} \textit{ JINST} \textbf{
  7} (2012) P10002,
  \href{http://dx.doi.org/10.1088/1748-0221/7/10/P10002}{\doi{10.1088/1748-0221/7/10/P10002}},
\href{http://www.arXiv.org/abs/1206.4071}{\texttt{arXiv:1206.4071}}.

\bibitem{Cacciari:2008gp}
\hrefCMSnoop {}{M.~Cacciari, G.~P. Salam, and G.~Soyez, ``The anti-$k_t$ jet
  clustering algorithm'',} \textit{ JHEP} \textbf{ 04} (2008) 063,
  \href{http://dx.doi.org/10.1088/1126-6708/2008/04/063}{\doi{10.1088/1126-6708/2008/04/063}},
  \href{http://www.arXiv.org/abs/0802.1189}{\texttt{arXiv:0802.1189}}.

\bibitem{FastJet}
\hrefCMSnoop {}{M.~Cacciari, G.~P. Salam, and G.~Soyez, ``{FastJet} user
  manual'',} \textit{ Eur. Phys. J. C} \textbf{ 72} (2012) 1896,
  \href{http://dx.doi.org/10.1140/epjc/s10052-012-1896-2}{\doi{10.1140/epjc/s10052-012-1896-2}},
  \href{http://www.arXiv.org/abs/1111.6097}{\texttt{arXiv:1111.6097}}.

\bibitem{Chatrchyan:2011ds}
\hrefCMSnoop {}{{CMS Collaboration}, ``{Determination of jet energy calibration
  and transverse momentum resolution in CMS}'',} \textit{ JINST} \textbf{ 6}
  (2011) P11002,
  \href{http://dx.doi.org/10.1088/1748-0221/6/11/P11002}{\doi{10.1088/1748-0221/6/11/P11002}},
\href{http://www.arXiv.org/abs/1107.4277}{\texttt{arXiv:1107.4277}}.

\bibitem{CMS:2015csvv2}
\href {http://cdsweb.cern.ch/record/2138504}{{CMS Collaboration},
  ``{Identification of b quark jets at the CMS Experiment in the LHC Run 2}'',}
  CMS Physics Analysis Summary CMS-PAS-BTV-15-001, 2016.

\bibitem{CMS:2012llxs}
\hrefCMSnoop {}{{CMS Collaboration}, ``Measurement of the \ttbar\ production
  cross section in the dilepton channel in pp collisions at {$\sqrt{s}
  =7$\TeV}'',} \textit{ JHEP} \textbf{ 11} (2012) 067,
  \href{http://dx.doi.org/10.1007/JHEP11(2012)067}{\doi{10.1007/JHEP11(2012)067}},
\href{http://www.arXiv.org/abs/1208.2671}{\texttt{arXiv:1208.2671}}.

\bibitem{CMS:2011WZ}
\hrefCMSnoop {}{{CMS Collaboration}, ``Measurements of inclusive {W} and {Z}
  cross sections in pp collisions at {$\sqrt{s} = 7$\TeV}'',} \textit{ JHEP}
  \textbf{ 01} (2011) 080,
  \href{http://dx.doi.org/10.1007/JHEP01(2011)080}{\doi{10.1007/JHEP01(2011)080}},
\href{http://www.arXiv.org/abs/1012.2466}{\texttt{arXiv:1012.2466}}.

\bibitem{CMS:2013b}
\hrefCMSnoop {}{{CMS Collaboration}, ``Identification of b-quark jets with the
  {CMS} experiment'',} \textit{ JINST} \textbf{ 8} (2013) P04013,
  \href{http://dx.doi.org/10.1088/1748-0221/8/04/P04013}{\doi{10.1088/1748-0221/8/04/P04013}},
\href{http://www.arXiv.org/abs/1211.4462}{\texttt{arXiv:1211.4462}}.

\bibitem{Khachatryan:2014qaa}
\hrefCMSnoop {}{{CMS Collaboration}, ``Search for the associated production of
  the {H}iggs boson with a top-quark pair'',} \textit{ JHEP} \textbf{ 09}
  (2014) 087,
  \href{http://dx.doi.org/10.1007/JHEP09(2014)087}{\doi{10.1007/JHEP09(2014)087}},
  \href{http://www.arXiv.org/abs/1408.1682}{\texttt{arXiv:1408.1682}}.
[Erratum: \DOI{10.1007/JHEP10(2014)106}].

\bibitem{CMS:2016ttbb}
\hrefCMSnoop {}{{CMS Collaboration}, ``{Measurement of the cross section ratio
  $\sigma_\mathrm{t \bar{t} b \bar{b}} / \sigma_\mathrm{t \bar{t} jj }$ in pp
  collisions at $\sqrt{s}$ = 8 TeV}'',} \textit{ Phys. Lett. B} \textbf{ 746}
  (2015) 132,
  \href{http://dx.doi.org/10.1016/j.physletb.2015.04.060}{\doi{10.1016/j.physletb.2015.04.060}},
\href{http://www.arXiv.org/abs/1411.5621}{\texttt{arXiv:1411.5621}}.

\bibitem{bdt1}
L.~Breiman, J.~Friedman, R.~A. Olshen, and C.~J. Stone, ``{Classification and
  regression trees}''.
\newblock {Chapman and Hall/CRC}, 1984.
\newblock
ISBN~0412048418, 9780412048418,

\bibitem{bdt2}
\hrefCMSnoop {}{H.~J. Friedman, ``Recent advances in predictive (machine)
  learning'',} \textit{ J. Classif.} \textbf{ 23} (2006) 175,
  \href{http://dx.doi.org/10.1007/s00357-006-0012-4}{\doi{10.1007/s00357-006-0012-4}}.

\bibitem{tmva}
\href {http://pos.sissa.it/archive/conferences/050/040/ACAT_040.pdf}{H.~Voss,
  A.~H{\"o}cker, J.~Stelzer, and F.~Tegenfeldt, ``{TMVA}, the Toolkit for
  Multivariate Data Analysis with {ROOT}'',} in \textit{ XIth International
  Workshop on Advanced Computing and Analysis Techniques in Physics Research
  (ACAT)}, p.~40.
\newblock 2007.
\newblock
\href{http://www.arXiv.org/abs/physics/0703039}{\texttt{arXiv:physics/0703039}}.
\newblock

\bibitem{sphericity}
\hrefCMSnoop {}{J.~D. Bjorken and S.~J. Brodsky, ``{Statistical model for
  electron-positron annihilation into hadrons}'',} \textit{ Phys. Rev. D}
  \textbf{ 1} (1970) 1416,
\href{http://dx.doi.org/10.1103/PhysRevD.1.1416}{\doi{10.1103/PhysRevD.1.1416}}.

\bibitem{CMS-PAS-LUM-15-001}
\href {http://cdsweb.cern.ch/record/2138682}{{CMS Collaboration}, ``{CMS
  luminosity measurement for the 2015 data-taking period}'',} CMS Physics
  Analysis Summary CMS-PAS-LUM-15-001, 2016.

\bibitem{ttNNLO}
\hrefCMSnoop {}{M.~Czakon, P.~Fiedler, and A.~Mitov, ``Total Top-Quark
  Pair-Production Cross Section at Hadron Colliders through
  {$O(\alpha^4_S)$}'',} \textit{ Phys. Rev. Lett.} \textbf{ 110} (2013) 252004,
  \href{http://dx.doi.org/10.1103/PhysRevLett.110.252004}{\doi{10.1103/PhysRevLett.110.252004}},
\href{http://www.arXiv.org/abs/1303.6254}{\texttt{arXiv:1303.6254}}.

\bibitem{Khachatryan:2016kdb}
\hrefCMSnoop {}{{CMS Collaboration}, ``{Jet energy scale and resolution in the
  CMS experiment in pp collisions at 8 TeV}'',} \textit{ JINST} \textbf{ 12}
  (2017) P02014,
  \href{http://dx.doi.org/10.1088/1748-0221/12/02/P02014}{\doi{10.1088/1748-0221/12/02/P02014}},
\href{http://www.arXiv.org/abs/1607.03663}{\texttt{arXiv:1607.03663}}.

\bibitem{RooStats:2010}
L.~Moneta\href
  {https://pos.sissa.it/archive/conferences/093/057/ACAT2010_057.pdf}{
  {et~al.}, ``{The RooStats Project}'',} in \textit{ 13th International
  Workshop on Advanced Computing and Analysis Techniques in Physics Research {
  (ACAT2010)}}, volume ACAT2010, p.~057.
\newblock 2010.
\newblock
\href{http://www.arXiv.org/abs/1009.1003}{\texttt{arXiv:1009.1003}}.
\newblock

\bibitem{Read:2002hq}
\hrefCMSnoop {}{A.~L. Read, ``Presentation of search results: the {$CL_{s}$}
  technique'',} \textit{ J. Phys. G} \textbf{ 28} (2002) 2693,
\href{http://dx.doi.org/10.1088/0954-3899/28/20/313}{\doi{10.1088/0954-3899/28/20/313}}.

\bibitem{Junk:1999kv}
\hrefCMSnoop {}{T.~Junk, ``Confidence level computation for combining searches
  with small statistics'',} \textit{ Nucl. Instrum. Meth. A} \textbf{ 434}
  (1999) 435,
  \href{http://dx.doi.org/10.1016/S0168-9002(99)00498-2}{\doi{10.1016/S0168-9002(99)00498-2}},
\href{http://www.arXiv.org/abs/hep-ex/9902006}{\texttt{arXiv:hep-ex/9902006}}.

\bibitem{Cowan:2011js}
\hrefCMSnoop {}{G.~Cowan, K.~Cranmer, E.~Gross, and O.~Vitells, ``{Asymptotic
  formulae for likelihood-based tests of new physics}'',} \textit{ Eur. Phys.
  J. C.} \textbf{ 71} (2011) 1554,
  \href{http://dx.doi.org/10.1140/epjc/s10052-011-1554-0}{\doi{10.1140/epjc/s10052-011-1554-0}},
\href{http://www.arXiv.org/abs/1007.1727}{\texttt{arXiv:1007.1727}}.

\bibitem{CMS-NOTE-2011-005}
\href {https://cds.cern.ch/record/1379837}{{The ATLAS and CMS Collaborations
  and the LHC Higgs Combination Group}, ``{Procedure for the LHC Higgs boson
  search combination in Summer 2011}'',} (2011).

\end{thebibliography}\endgroup

\cleardoublepage \appendix\section{The CMS Collaboration \label{app:collab}}\begin{sloppypar}\hyphenpenalty=5000\widowpenalty=500\clubpenalty=5000\textbf{Yerevan Physics Institute,  Yerevan,  Armenia}\\*[0pt]
A.M.~Sirunyan, A.~Tumasyan
\vskip\cmsinstskip
\textbf{Institut f\"{u}r Hochenergiephysik,  Wien,  Austria}\\*[0pt]
W.~Adam, E.~Asilar, T.~Bergauer, J.~Brandstetter, E.~Brondolin, M.~Dragicevic, J.~Er\"{o}, M.~Flechl, M.~Friedl, R.~Fr\"{u}hwirth\cmsAuthorMark{1}, V.M.~Ghete, C.~Hartl, N.~H\"{o}rmann, J.~Hrubec, M.~Jeitler\cmsAuthorMark{1}, A.~K\"{o}nig, I.~Kr\"{a}tschmer, D.~Liko, T.~Matsushita, I.~Mikulec, D.~Rabady, N.~Rad, B.~Rahbaran, H.~Rohringer, J.~Schieck\cmsAuthorMark{1}, J.~Strauss, W.~Waltenberger, C.-E.~Wulz\cmsAuthorMark{1}
% ...
\vskip\cmsinstskip
% ...
1:~~Also at Vienna University of Technology, Vienna, Austria\\
% ...

\end{sloppypar}
\end{document}